\documentclass[amsmath,amssymb,nofootinbib,notitlepage,superscriptaddress,twocolumn]{revtex4-1}
\pdfoutput=1

\usepackage{aas_macros,esint,overpic}
\usepackage[bottom]{footmisc}
\usepackage[colorlinks=true]{hyperref}
\usepackage[raggedright]{subfigure}
\let\originalleft\left
\let\originalright\right
\renewcommand{\left}{\mathopen{}\mathclose\bgroup\originalleft}
\renewcommand{\right}{\aftergroup\egroup\originalright}

\DeclareMathOperator\sign{sign}

\begin{document}

\title{Measuring the shape of a black hole photon ring}

\author{Samuel E. Gralla}
\email{sgralla@email.arizona.edu}
\affiliation{Department of Physics, University of Arizona, Tucson, Arizona 85721, USA}

\begin{abstract}

General relativity predicts that gravitational lensing near black holes will produce narrow ``photon rings'' on images.   Building on recent work of Johnson, Lupsasca et al. focusing on circular rings, I calculate the long-baseline interferometric signature of a narrow feature of arbitrary shape and intensity.  The shape information is contained in the oscillation frequencies of the complex visibility as a function of baseline length, for each baseline angle.  These results raise the possibility of measuring the detailed shape of a photon ring and comparing to the precise predictions of general relativity.

\end{abstract}

\maketitle

\section{Introduction}

General relativity predicts that emission near black holes will be lensed into a series of increasingly narrow ``photon rings''  converging to a critical curve on the image plane  \cite{bardeen1973,luminet1979,beckwith-done2005,gralla-holz-wald2019,johnson-etal2020,gralla-lupsasca2020}.  Recently it was shown \cite{johnson-etal2020} that a circular ring of diameter $d$ and width $w\ll d$ has a universal interferometric signature,
\begin{align}\label{johnson}
    V(u,\varphi) \approx \frac{a(\varphi) \cos(\pi d u) + b(\varphi) \sin(\pi d u)}{\sqrt{u}},
\end{align}
holding on baseline lengths $u$ in the range
\begin{align}\label{regime}
    \frac{1}{d} \ll u \ll \frac{1}{w}.
\end{align}
Here $V$ is the complex visibility as a function of polar coordinates $(u,\varphi)$, and the complex coefficients $a$ and $b$ are related to the intensity profile around the ring.  This is an exciting result because the ring diameter (a property of the black hole) is encoded in the periodicity $2/d$, cleanly separated from emission profile information contained in $a$ and $b$.  The distinctive $2/d$ periodicity is unlikely to be contaminated by other sources of emission relevant on these baselines, which should be transient or stochastic in nature.  Thus one can potentially detect a black hole photon ring via the periodicity of an observed visibility.

An even more exciting prospect would be to use this approach for precision measurements of black hole properties as well as tests of general relativity.  However, the ring predicted by general relativity is not precisely circular, and the small deviations encode information about mass, spin, and inclination \cite{johanssen-psaltis2010}.  Furthermore, null tests of the theory will require understanding the signature of more general curve shapes.  Thus Eq.~\eqref{johnson} must be generalized beyond circular rings.  Ref.~\cite{johnson-etal2020} suggested that in general an angle-dependent periodicity would encode some kind of angle-dependent diameter, but did not  give details.  In this paper we will derive the full, detailed universal interferometric signature of a thin curve of arbitrary shape and intensity, laying a theoretical foundation capable in principle of precision black hole measurements and tests of general relativity.

\begin{figure*}
    \centering
    \includegraphics[scale=.25]{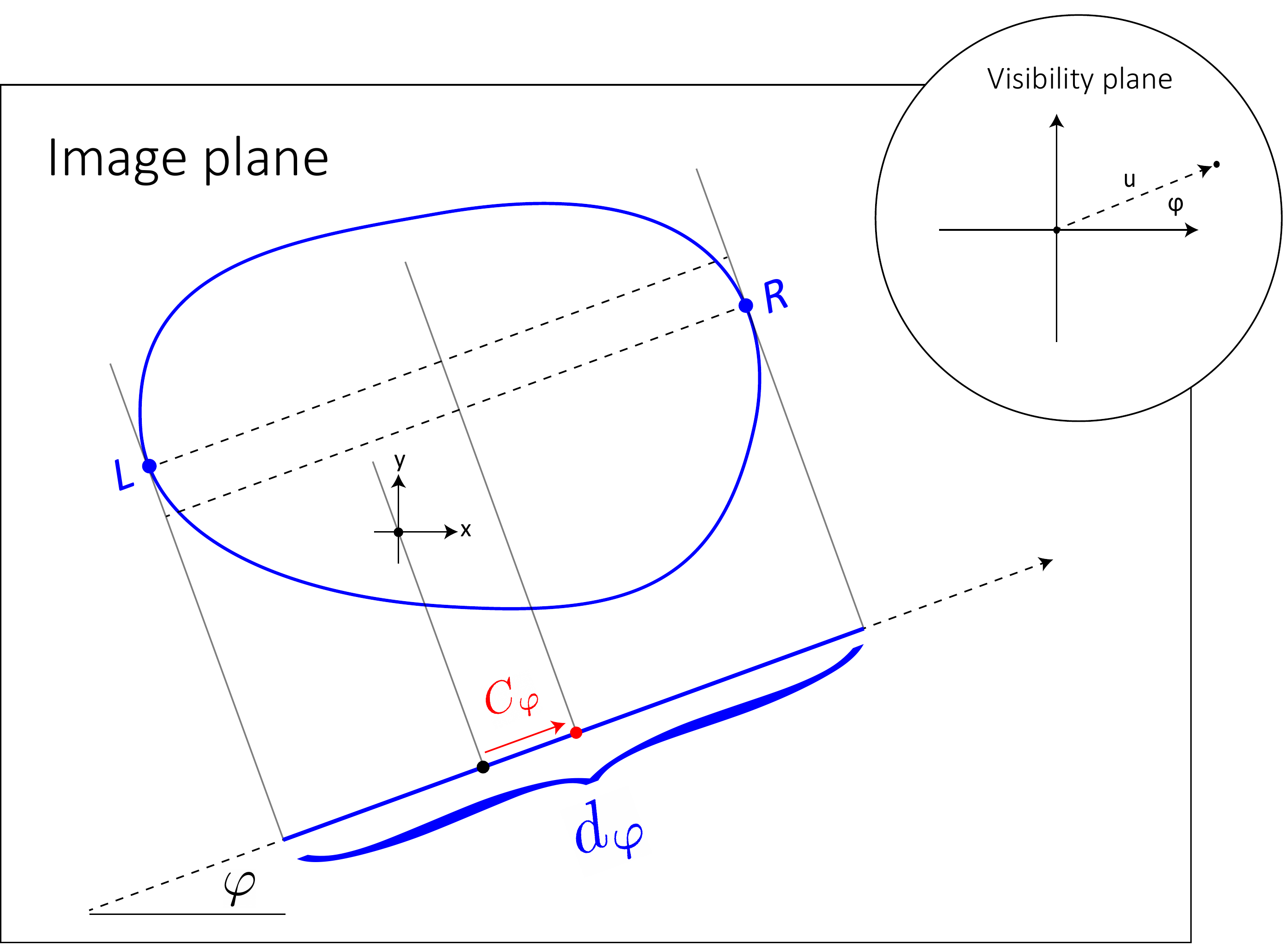} \qquad 
    \includegraphics[scale=.82]{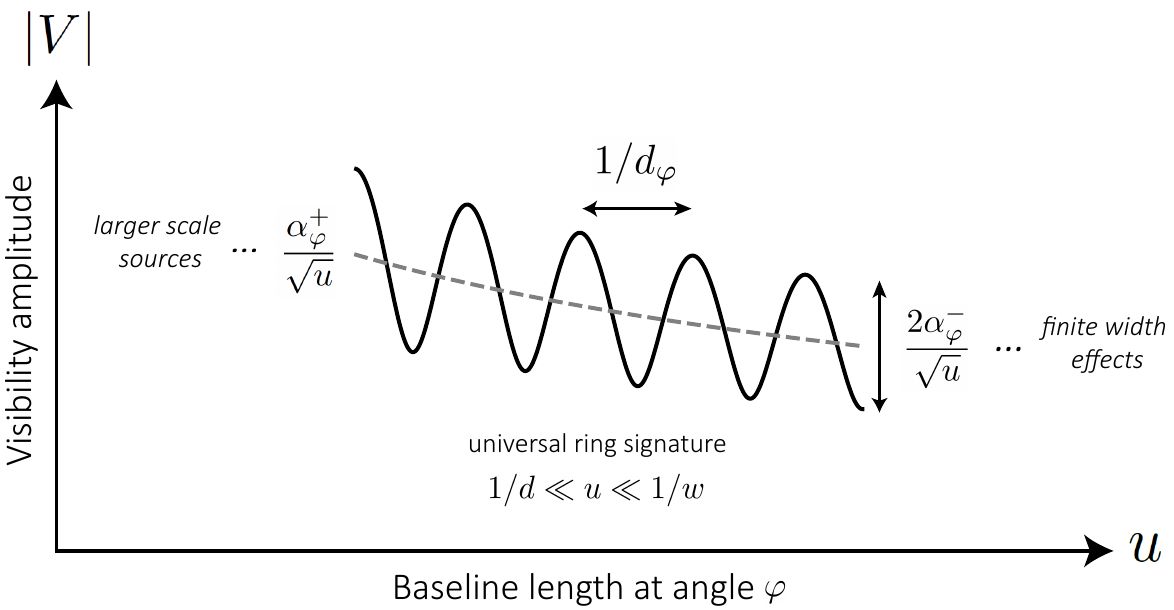}
    \caption{A sky intensity narrowly peaked around a closed convex curve of typical diameter $d$ and width $w$ has a universal interferometric signature on baselines $1/d \ll u \ll 1/w$.  At each angle $\varphi$, the periodicity $1/d_\varphi$ in the visibility amplitude encodes the projected diameter $d_\varphi$, while the maximum and minimum values encode the intensity at the edge points $L$ and $R$.  The projected centroid $C_\varphi$ is contained in the full complex visibility.}
    \label{fig:mic drop}
\end{figure*}

Fig.~\ref{fig:mic drop} left shows the geometric ingredients relevant to the main result, restricting for simplicity to closed, convex curves. The measurable diameter is the \textit{projected} diameter $d_\varphi$, which is equivalent to the length of the one-dimensional shadow cast by the curve when a light source shines on the baseline.\footnote{In other words, long baselines see the shadow of the photon ring!}  The visibility contains intensity information from the two edge points $L$ and $R$ (and only these points!) as well as an overall phase encoding the projected centroid position $C_\varphi$.  

To present the result, let $\mathcal{I}$ denote the local, transverse-integrated intensity of the source curve, such that the total flux density is $\int \mathcal{I}(s) ds$, with $s$ the arclength.  This $\mathcal{I}$ has units of intensity times angle, and most properly would be called the linear angular density of flux density.  (We will stick with ``integrated intensity''.)  Then the visibility of a closed, convex curve in the regime \eqref{regime} is given by
\begin{align}\label{mylove}
    V & \approx \frac{e^{-2 \pi i C_\varphi u}}{\sqrt{u}} \left(\alpha^L_\varphi \ \! e^{-\frac{i\pi}{4}}e^{ i \pi d_\varphi u} + \alpha^R_\varphi \ \! e^{\frac{i\pi}{4}} e^{- i \pi d_\varphi u} \right),
\end{align}
where the real, positive coefficients  $\alpha^L_\varphi$ and $\alpha^R_\varphi$ are determined from the integrated intensity $\mathcal{I}$ and radius of curvature $\mathcal{R}$ at the edge points by
\begin{align}\label{alpha}
    \alpha^{(L/R)}_\varphi = \left.\left(\mathcal{I} \sqrt{\mathcal{R}}\right)\right|_{L/R}.
\end{align}
The generalization to arbitrary smooth curves takes a similar form (Eq.~\eqref{bowerbird} below), with contributions from all points where the baseline direction is perpendicular to the curve.  In Ref.~\cite{gralla-lupsasca2020b} we give an explicit formula for reconstructing the original curve from its signature \eqref{mylove} or \eqref{bowerbird}.  One can then calculate $\mathcal{R}$ and use it to infer the intensity $\mathcal{I}$ (see Eqs.~\eqref{alpha} and \eqref{bowerbird}).  That is, the universal interferometric signature contains the full information about curve shape and intensity.

The complex visibility \eqref{mylove} has two fundamental frequencies $C_\varphi\pm d_\varphi$ encoding the projected diameter and centroid.  The visibility \textit{amplitude} encodes the diameter alone as
\begin{align}\label{Visamp}
    |V| & \approx \frac{1}{\sqrt{u}} \sqrt{ (\alpha^L_\varphi)^2 + (\alpha^R_\varphi)^2  + 2 \alpha^L_\varphi \alpha^R_\varphi \sin (2\pi d_\varphi u)}.
\end{align}
Apart from the slow $1/\sqrt{u}$ decay, this function has periodicity $1/d_\varphi$, alternating every half-period  between minima and maxima given by
\begin{align}
    V_{\rm max} = \frac{\alpha^L_\varphi + \alpha^R_\varphi}{\sqrt{u}}, \qquad V_{\rm min} = \frac{|\alpha^L_\varphi - \alpha^R_\varphi|}{\sqrt{u}}.
\end{align}
If $\alpha^\pm_\varphi$ denote the larger $(+)$ and smaller $(-)$ of $\alpha^L_\varphi$ and $\alpha^R_\varphi$, then the  average value $\bar{V}=(V_{\rm max}+V_{\rm min})/2$ and oscillation height $\Delta V=V_{\rm max}-V_{\rm min}$ are simply
\begin{align}\label{oscillate}
    \bar{V} = \frac{\alpha^+_\varphi}{\sqrt{u}}, \qquad \Delta V = \frac{2\alpha^-_\varphi}{\sqrt{u}}.
\end{align}
These properties are illustrated in Fig.~\ref{fig:mic drop} right.  

From a measurement of the visibility amplitude alone, one may infer the set of projected diameters $d_\varphi$ of the curve.  This is enough to reconstruct the ``hull'' of the shape, which encodes its rough overall dimensions \cite{gralla-lupsasca2020b}.

These results provide a direct map between the source properties and the observed visibility.  It is notable that this mapping is \textit{local},\footnote{The apparently non-local formulas for $a(\varphi)$ and $b(\varphi)$ given in Eq.~(20) of Ref.~\cite{johnson-etal2020} (involving an infinite sum over multipoles) are in fact local and agree with the general result \eqref{mylove}.}   in the sense that each baseline angle $\varphi$ provides information about a discrete set of points on the image plane (denoted $L$ and $R$ for closed convex curves).  This is somewhat surprising in light of the general tendency of Fourier transform to smear out information, and has its origin in the singular nature of the projection at the special points.  The locality may have important practical consequences.  For example, if only a portion of the angles $\varphi$ can be observed due to experimental limitations, one can still perform detailed measurements of the corresponding portions of the curve.  Similarly, if part of the photon ring is obscured, the resulting open curve can still be probed via the more general result \eqref{bowerbird}.

The appearance of the radius of curvature $\mathcal{R}$ in the coefficients \eqref{alpha} is worthy of further discussion.   Although the photon ring predicted by general relativity is always nearly circular in terms of effective diameter \cite{bardeen1973,chan-psaltis-ozel2013}, it can become highly non-circular in terms of radius of curvature, since $\mathcal{R} \to \infty$ over a large portion (the ``NHEKline'') in the extremal limit \cite{bardeen1973,gralla-lupsasca-strominger2018}.  The interferometric signature of the NHEKline is thus enhanced relative to other portions of the curve with comparable intensity.

In the remainder of the paper we derive these results, first in the axisymmetric case and then in general.

\section{Axisymmetric case}\label{sec:axi}

The complex visibility is defined to be the Fourier transform of the sky intensity $I(\vec{r})$,
\begin{align}
    V(\vec{u}) = \int I(\vec{r}) e^{- 2\pi i \vec{u} \cdot \vec{r}} d^2 \! \vec{r}.
\end{align}
Here $\vec{r}=(x,y)$ is the sky position in radians, while $\vec{u}=(u_x,u_y)$ is the baseline in units of observation wavelength.  We will use polar coordinates  $(u,\varphi)$ for the visibility plane.

Consider a perfectly circular ring with uniform intensity.  For $u \ll 1/w$ we cannot resolve the width, and hence the intensity profile may be approximated as a delta function,
\begin{align}\label{Iaxi}
    I = \mathcal{I} \delta(r-r_0),
\end{align}
where $r=\sqrt{x^2+y^2}$ and $r_0$ is the ring radius.  The coefficient $\mathcal{I}$ is the ``integrated intensity'' defined above Eq.~\eqref{mylove}, such that the total flux density is $2\pi r_0 \mathcal{I}$.  The Fourier transform of any axisymmetric profile is the order-zero Hankel transform,
\begin{align}\label{universal}
    V(u) = 2\pi \int I(r) J_0(2\pi r u) r dr,
\end{align}
where $J_n$ denotes the Bessel function of the first kind.  Our  delta-function profile \eqref{Iaxi} therefore gives just
\begin{align}\label{VBessel}
    V = 2\pi r_0 \mathcal{I} J_0(2 \pi r_0 u).
\end{align}
Using the large-argument approximation for the Bessel function then gives
\begin{align}\label{Vaxi}
 V \approx 2 \mathcal{I}\sqrt{r_0} \frac{\cos\left(2 \pi r_0 u-\frac{\pi}{4} \right) }{\sqrt{u}},
 \end{align}
 holding in the regime \eqref{regime}.  Although this formula could be elegantly written in terms of the total flux density  $V_0=2\pi r_0 \mathcal{I}$ and the large parameter $2 \pi r_0 u$, this form does not generalize well to the non-axisymmetric case.

The Hankel transform approach is the easiest path to the answer, but it provides little insight into the nature of the phenomenon and does not easily generalize to non-circular rings.  Let us instead view the problem using the projection-slice theorem, which states that the visibility at angle $\varphi$ is given by the one-dimensional Fourier transform of the projection of the intensity onto a line at angle $\varphi$ in the image plane.  We will consider the $x$-axis ($\varphi=0$) without loss of generality.  Expressing the $\delta$-function in terms of $y$ as
\begin{align}
    \delta(r-r_0) = \sum_{+,-} 
    \frac{\delta\left( y \pm \sqrt{r_0^2 - x^2} \right)}{\sqrt{1-(x/r_0)^2}},
\end{align}
the projection is just
\begin{align}\label{axiproj}
    \mathcal{P}_x I = \int I dy = 2 r_0 \mathcal{I} \frac{H(r_0^2-x^2)}{\sqrt{r_0^2-x^2}},
\end{align}
where $H$ is the Heaviside step function.  The visibility is given by the Fourier transform $\mathcal{F}_x$,
\begin{align}\label{thing}
    V = \mathcal{F}_x \mathcal{P}_x I & = 2 r_0 \mathcal{I} \int \frac{H(r_0^2-x^2)}{\sqrt{r_0^2-x^2}} e^{- 2\pi i x u} dx,
\end{align}
which indeed reproduces \eqref{VBessel} after sufficient perusal of integral tables.

This approach still relies on an obscure Fourier transform, but we are now closer to the heart of the phenomenon.  The key observation is that the projection $\mathcal{P}_xI$ being Fourier transformed has singularities at $x=\pm r_0$.  While the full function $\mathcal{P}_xI$ is needed to recover the full Bessel function visibility \eqref{VBessel}, only these singularities are necessary to recover large-argument limit \eqref{Vaxi} of interest.  The leading singular behavior of \eqref{axiproj} is given by
\begin{align}
    (\mathcal{P}_x I)_{\rm sing} = \sqrt{2 r_0} \mathcal{I} \left(\frac{H(r_0+x)}{\sqrt{r_0+x}} +  \frac{H(r_0-x)}{\sqrt{r_0-x}} \right).
\end{align}
The Fourier transform of this function is easily computed using the elementary transform
\begin{align}\label{elementary}
    \int \frac{H(z)}{\sqrt{z}} e^{-2\pi i k z}dz = \frac{e^{-\frac{i \pi}{4}\sign k}}{\sqrt{2|k|}},
\end{align}
(holding for real $k \neq 0$), giving
\begin{align}
    \mathcal{F}_x (\mathcal{P}_x I)_{\rm sing} & = \sqrt{2 r_0} \mathcal{I} \left( \frac{e^{ 2\pi i r_0 u - i \pi/4}}{\sqrt{2  u}} +  \frac{e^{ -2\pi i r_0 u + i \pi/4}}{\sqrt{2 u}} \right) \nonumber \\
    & = 2 \mathcal{I} \sqrt{r_0}\frac{\cos\left(2\pi r_0 u-\frac{\pi}{4}\right)}{\sqrt{u}},
\end{align}
in agreement with Eq.~\eqref{Vaxi}.

Let us summarize.  The projection of the ring has singularities at the projected edges, which dominate the Fourier transform at large baselines.  Each singularity contributes a term decaying like $1/\sqrt{u}$, with a phase oscillation related to its coordinate position.  The resulting visibility oscillates at the rate determined by the difference in coordinate positions, i.e., the diameter of the ring.  These properties will generalize straightforwardly to general curve shapes, as we now show.

\section{General derivation}\label{sec:gen}

Suppose that the narrow intensity profile takes the shape of a smooth plane curve.  Parameterizing the curve as  $\vec{r}_0(s)=(x_0(s),y_0(s))$ using the arclength $s$, the intensity is given by
\begin{align}
    I = \int \mathcal{I}(s) \delta(x-x_0(s))\delta(y-y_0(s)) ds,
\end{align}
where the integrated intensity $\mathcal{I}(s)$ now varies around the curve.  Let us again choose the $x$-axis for the projection,
\begin{align}\label{eek}
    \mathcal{P}_x I & = \int I dy = 
    \int \mathcal{I}(s) \delta(x-x_0(s)) ds.
\end{align}
This integral defines a function of $x$.  At a given value of $x$ there may zero, one, or more segments of the curve being integrated over.  We may imagine plotting the curve, drawing a vertical line at $x$, and counting the number $N$ of intersections.  As long as the vertical line is never tangent to the curve, we may write
\begin{align}\label{sum}
    \delta(x-x_0(s))= \sum_{i=1}^{N} \frac{\delta(s-s_i)}{|x_0'(s_i)|},
\end{align}
where $s_i$ are the parameter values of the intersections at $x$ (i.e., $x_0(s_i)=x$).  The integral \eqref{eek} is then just
\begin{align}\label{yikes}
\mathcal{P}_x I (x) = \sum_{\rm branches} \frac{\mathcal{I}}{|dx_0/ds|},
\end{align}
where the sum ranges over all intersections present at $x$.  

Eq.~\eqref{yikes} does not hold pointwise at points $x$ where the vertical line intersects the curve at a point of tangency ($dx_0/ds=0$), but we shall see that it defines a unique distribution over $x$, to which the theory of Fourier transform straightforwardly applies.  To see this we must analyze the nature of the singularities at vertical points $(dx_0/ds=0)$, which depends on the behavior of higher derivatives of the function $x_0(s)$.  

First consider the ``generic'' case where the second derivative is non-vanishing and the curve does not end at the vertical point.  Near the parameter value $\tilde{s}$ of the vertical point, we have
\begin{align}
    x_0 = \tilde{x} + \frac{1}{2} \tilde{Q} (s-\tilde{s})^2 + O(s-\tilde{s})^3, 
\end{align}
where
\begin{align}
    \tilde{x} = x_0(\tilde{s}), \qquad \tilde{Q} = x_0''(\tilde{s}) \neq 0.
\end{align}
In this region we have 
\begin{align}\label{baby pony}
    \left|\frac{dx_0}{ds}\right| \approx |\tilde{Q}(s-\tilde{s})| \approx \sqrt{2\tilde{Q}(x_0-\tilde{x})},
\end{align}
where the quantity under the square root is positive by construction as $x_0 \to \tilde{x}$.  

Now consider the contribution to the projected intensity \eqref{yikes} from near the vertical point $\tilde{s}$.  First suppose that $\tilde{Q}$ is positive, so that the curves continue to the right of the vertical point $(x>\tilde{x})$, but not to the left (like the letter ``C'').  From \eqref{yikes} and \eqref{baby pony}, the leading contribution of these curves in a small two-sided neighborhood of $x$ is 
\begin{align}\label{factorOf2}
     (\mathcal{P}_x I)_{\rm near \ \tilde{s}} = 2\tilde{\mathcal{I}}\frac{ H(x-\tilde{x})}{\sqrt{2\tilde{Q}(x-\tilde{x})}},
\end{align}
where $\tilde{\mathcal{I}}=\mathcal{I}(\tilde{s})$ is the integrated intensity at the vertical point.  If instead we have $\tilde{Q}<0$, then the curves are to the left and we require $H(\tilde{x}-x)$ instead.  We can represent both cases by including a factor of $\tilde{Q}$ in the Heaviside function, writing
\begin{align}
    (\mathcal{P}_x I)_{\rm near \ \tilde{s}} = 2 \tilde{\mathcal{I}}\frac{ H(\tilde{Q}(x-\tilde{x}))}{\sqrt{2\tilde{Q}(x-\tilde{x})}}.
\end{align}  
This is the leading contribution to the projected intensity from near a generic vertical point.  If all vertical points are generic, then the complete leading singular behavior of the projection is
\begin{align}\label{stallion}
(\mathcal{P}_x I)_{\rm sing} = \sum_{I} \sqrt{2}\mathcal{I}_I \frac{H(Q_I(x-x_I))}{\sqrt{Q_I(x-x_I)}},
\end{align}
where subscript $I$ indicates evaluation at the $I^{\rm th}$ vertical point.  We may now Fourier transform using Eq.~\eqref{elementary} to get the leading behavior on long baselines
\begin{align}
    V \approx \sum_I \frac{\mathcal{I}_I}{\sqrt{|Q_I|}} e^{- \frac{i\pi}{4} S_I} \frac{e^{-2\pi i x_I u}}{\sqrt{u}},\label{vermilion flycatcher}
\end{align}
with
\begin{align}
    S_I = \sign Q_I.
\end{align}
We write $\approx$ since Eq.~\eqref{vermilion flycatcher} holds in the regime \eqref{regime}, with $d$ a typical scale for the curve.

Finally we will rewrite this result in geometric language intrinsic to the curve.  Our projection was along the $x$-axis, but since the curve was described without any special choice of frame, we have lost no generality in making this choice.  To write the description in terms of an arbitrary baseline angle $\varphi$, let $\hat{u}$ denote the unit vector along the baseline.  (This vector is a ``radial vector'' in the visibility plane, but is viewed as constant in the image plane, pointing along the visibility direction $\varphi$ under consideration.)   Then we may rewrite $x_I$ and $Q_I$ as the invariant expressions
\begin{align}
    x_I & = \left(\vec{r}_0\cdot \hat{u} \right)_I \\ \qquad Q_I & = \left( \frac{d ^2 \vec{r}_0}{d s^2} \cdot \hat{u} \right)_I.\label{QI}
\end{align}
It will be helpful to change the names to emphasize that these quantities are now invariantly defined.  First, we will write $x_I=z_I$ for the projected distance, to avoid confusion with the coordinate axis $x$ in any given image description.  Second, we will re-express $Q_I$ in terms of the standard geometric notions for plane curves.  Recall that at any point along a plane curve, the unit normal $\hat{n}$ pointing toward the center of curvature is given by 
\begin{align}
    \hat{n} = \frac{d ^2 \vec{r}_0}{d s^2} \mathcal{R}, 
\end{align}
where $\mathcal{R}>0$ is the radius of curvature.  At each  point $I$ around our curve, by definition the normal $\hat{n}$ is either aligned or anti-aligned with the baseline direction $\hat{u}$, and from \eqref{QI} $Q_I$ may be written
\begin{align}
    Q_I = \left(\mathcal{R}^{-1} \hat{n} \cdot \hat{u}\right)_I = S_I  \mathcal{R}_I^{-1},
\end{align}
showing that the sign $S_I$ is given by
\begin{align}
    S_I = \left(\hat{n} \cdot \hat{u}\right)_I = \pm 1.
\end{align}
Collecting everything together, we can express the final result \eqref{vermilion flycatcher} in a form valid for any projection axis as
\begin{align}\label{bowerbird}
    V(u,\varphi) \approx \sum_I \mathcal{I}_I \sqrt{\mathcal{R}_I} e^{-\frac{i\pi}{4}S_I} \frac{e^{-2\pi i z_I u}}{\sqrt{u}}.
\end{align}
Here the sum ranges over the ($\varphi$-dependent) list of curve points where the baseline direction $\hat{u}(\varphi)$ is normal to the curve.  For each such point, $\mathcal{I}_I$ is the integrated intensity, $\mathcal{R}_I$ is the radius of curvature, $z_I=(\vec{r}_0 \cdot \hat{u})_I$ is the projected position, and  $\mathcal{S}_I=(\hat{n} \cdot \hat{u})_I$ is $+1$ or $-1$ depending on whether the baseline direction $\hat{u}(\varphi)$ points toward or away from the center of curvature, respectively.

At each angle $\varphi$, the form \eqref{bowerbird} contains some given number of oscillation frequencies $z_I$ (the number of terms in the sum), determined by how many times the curve is normal to the the baseline direction.  Given a sufficiently accurate observation, the number of oscillation frequencies can be inferred from the signal, and the set of functionally independent parameters $\{z_I, \mathcal{I}\sqrt{\mathcal{R}_I}\}$ (for all $I$) can be determined.  As the baseline angle $\varphi$ is varied from $0$ to $\pi$, every point on the curve appears as a point $I$ in the sum \eqref{bowerbird} for some $\varphi$.  One can thus in principle determine $z$ and $\mathcal{I} \sqrt{R}$ everywhere on the curve.  Since a plane curve is specified locally by a single free function (such as $y(x)$), the determination of $z$ should be enough to reconstruct the complete curve shape.  I am grateful to A. Lupsasca for conversations on this matter, and we have prepared a joint publication including an explicit formula for reconstructing the curve \cite{gralla-lupsasca2020b}.   Once the curve has been constructed from $z$, the radius of curvature $\mathcal{R}$ can be calculated and the integrated intensity profile $\mathcal{I}$ can be determined from the measurement of $\mathcal{I}\sqrt{R}$.

In deriving Eq.~\eqref{bowerbird}, we assumed that none of the points $I$ are endpoints of the curve.  This can be true for all angles $\varphi$ only if the curve is closed (assuming, of course, that it occupies a finite region of the image plane).  However, the generalization to open curves is straightforward.  Tracing through the derivation, we see that a factor of $2$ appeared in Eq.~\eqref{factorOf2} to account for the continuation of the curve in both directions.  If the curve in fact ends at this point, the factor of 2 should be dropped.  Thus Eq.~\eqref{bowerbird} generalizes to open curves with the proviso that the summand should contain an extra factor of $1/2$ whenever the point $I$ is an endpoint of the curve. However, this affects only a measure-zero set of baseline angles $\varphi$, and can likely be ignored in practice.

Eq.~\eqref{bowerbird} also assumes that the radius of curvature $\mathcal{R}$ is finite at all points along the curve.  If the curve has points where $\mathcal{R}$ blows up, one must redo the analysis for those points.  Returning to the derivation, we see that $\tilde{Q}\neq 0$ was required to obtain Eq.~\eqref{baby pony}.  If $\tilde{Q}$ vanishes (infinite radius of curvature), then $dx_0/ds$ will instead scale as $|\tilde{x}-x_0|^{1-1/n}$, where $n$ is the order of the first non-vanishing derivative of $x_0(s)$.  Then by \eqref{yikes} the projection $\mathcal{P}_xI$ will scale as $(x-x_0)^{1/n-1}$, giving rise to a long-baseline visibility decaying as $u^{1/n}$ (with the same oscillations $e^{-2\pi i x_I u}$).  If there is a truly straight portion of the curve, such that all derivatives of $x_0(s)$ are vanishing, then from \eqref{eek} we see that $\mathcal{P}_x I$ will contain a delta function in $x$, giving a Fourier transform that is constant in magnitude (but still oscillates like $e^{-2\pi i x_I u}$).

That is, when the coefficient $\mathcal{R}_I$ in \eqref{bowerbird} becomes infinite, the true answer is generally a more slowly decaying function of $u$, approaching the extreme of non-decay when there is a precisely straight segment.  However, this behavior occurs only at a measure-zero set of baseline angles $\varphi$, and can likely be handled in practice by simply excluding a very small region of angles $\varphi$ near the offending direction.  The nearby visibility amplitude will increase on account of the $\sqrt{\mathcal{R}_I}$ prefactor in Eq.~\eqref{bowerbird}, without need to explicitly consider a slower falloff in $u$.

Finally, consider the special case of a closed, convex curve.  In this case there are precisely two points $I$ at every angle, the left and right points shown in Fig.~\ref{fig:mic drop} left.  These have signs $S_L=+1$ and $S_R=-1$, respectively, and \eqref{bowerbird} becomes 
\begin{align}
    V & \approx \mathcal{I}_L \sqrt{\mathcal{R}_L} e^{-\frac{i\pi}{4}} \frac{e^{-2\pi i z_L u}}{\sqrt{u}} + \mathcal{I}_R \sqrt{\mathcal{R}_R}  e^{\frac{i\pi}{4}} \frac{e^{-2\pi i z_R u}}{\sqrt{u}}. \nonumber
\end{align}
The expression \eqref{mylove} given in the introduction arises after introducing the projected diameter and centroid,
\begin{align}
    d_\varphi & = z_R - z_L \\
    C_\varphi & = \frac{1}{2} \left( z_L + z_R \right).
\end{align}

\section*{Acknowledgements}

This work was supported in part by NSF grant PHY-1752809 to the University of Arizona.

\bibliography{moon}
\bibliographystyle{utphys}

\end{document}